\documentclass[twocolumn,showpacs,preprintnumbers,amsmath,amssymb,pra]{revtex4}

\usepackage{color}
\usepackage{amsmath}
\usepackage{graphicx}
\usepackage{dcolumn}
\usepackage{bm}
\usepackage[amssymb]{SIunits}

\begin{document}

\title{Geometric phases in open tripod systems}
\author{Ditte M\o ller}
\email{dittem@phys.au.dk}
\author{Lars Bojer Madsen}
\author{Klaus M\o lmer}
\affiliation{Lundbeck Foundation Theoretical Center for Quantum System Research,\\
Department of Physics and Astronomy, University of Aarhus, DK-8000,
Denmark.}
\date{\today }

\begin{abstract}
We first consider stimulated Raman adibatic passages (STIRAP) in a
closed four-level tripod system. In this case, the adiabatic
eigenstates of the system acquire real geometric phases. When the
system is open and subject to decoherence they acquire complex
geometric phases that we determine by a Monte Carlo wave function
approach. We calculate the geometric phases and the state evolution
in the closed as well as in the open system cases and describe the
deviation between these in terms of the phases acquired. When the
system is closed, the adiabatic evolution implements a Hadamard
gate. The open system implements an imperfect gate and hence has a
fidelity below unity. We express this fidelity in terms of the
acquired geometric phases.
\end{abstract}


\maketitle

\section{Introduction}\label{sec:introduction}
Each eigenstate of a non-degenerate quantum system that evolves
adiabatically in time acquires a well-defined phase in addition to
the usual dynamic phase. The former is called the geometric phase
because it depends on the path traversed in Hilbert space
\cite{berry84}. With the current interest in quantum information and
computation, the geometric phase has received new attention because
it is expected to be robust against some sources of decoherence
\cite{ellinas01,solinas04,fuentes05,zhu05} and hence useful, e.g.,
for creation of quantum gates. To study its robustness under
decoherence it is essential to generalize the concept of geometric
phase to open quantum systems. Various proposals have been made
\cite{sjoqvist06,dasgupta07} and they all point to the problem that
phase information tends to be lost when the system is open and
decoheres, e.g., due to spontaneous emission.

The full system including decoherence can still be described, for
example, by the density matrix approach, which predicts the relative
phases between the involved basis states, but the information about
the phases acquired by each eigenstate is not available nor is the
information about the geometric or dynamic nature of the phase. In
this work we are interested in the phase dynamics of the states of
the physical system when the system is subject to decoherence. To
this end we consider the Monte Carlo wave function (MCWF) approach
\cite{molmer93,molmer96}. We follow each wave function trajectory
and calculate the complex geometric phases that are acquired by the
adiabatic eigenstates. The MCWF approach has the advantage that we
gain information about the evolution of the single trajectories and
the dynamic or geometric nature of the phases. Not surprisingly, on
average the trajectories reproduce the density matrix result. We
show how the evolution of the open system can be described by
adiabatic eigenstates that acquire complex geometric phases and we
discuss the deviation from the closed system case in terms of the
phases acquired. We consider a tripod system with three laser fields
applied and present a full calculation of the adiabatic evolution of
the wave function of the system. The adiabatic evolution considered
consist of a sequence of stimulated Raman adiabatic passages
(STIRAP)\cite{bergmann98} and implements the Hadamard gate. We
quantify the effect of decoherence in terms of the fidelity of this
gate and identify the role of the complex geometric phases.

The paper is organized as follows. In Sec. \ref{sec:background} we
review the theory of geometric phases, we explain their appearance
in STIRAP processes and present the tripod system. In Sec.
\ref{sec:closedsystem} we present calculations for the closed system
and in Sec. \ref{sec:opensystem} we derive the full solution for the
evolution of the open system evolution and calculate the fidelity of
the Hadamard gate. Sec. \ref{sec:conclusion} concludes.

\section{Background}\label{sec:background}
\subsection{Geometric phase}
The adiabatic theorem states that for a given set of instantaneous
eigenstates, $\psi_n(t)$, and eigenenergies, $E_n(t)$, of a
timedependent Hamiltonian, there is no population transfer between
the eigenstates if these vary slowly compared with the energy
difference between eigenstates \cite{ messiah61}
\begin{equation}
\left|\frac{\partial\psi_n}{\partial
t}\right|\ll\frac{|E_n-E_m|}{\hbar}.
\end{equation}
In this adiabatic regime the eigenstates do not only acquire a
dynamic phase, $\vartheta_n=\int^{t_f}_{t_i}E_n(t')/\hbar \textrm{ }
dt'$, but also a geometric phase, $\gamma_n$
\begin{equation}
\Psi(0)=\sum_n c_n(0)\psi_n(0) \xrightarrow{ad} \Psi(t)=\sum_n
c_n(0)e^{i(\vartheta_n+\gamma_n)}\psi_n(t),
\end{equation}
where the geometric part of the phase can be calculated directly
from the eigenstates \cite{berry84},
\begin{equation}
\gamma_{n}=i\int_{\bar{R}_i}^{\bar{R}_f}\langle \psi_n(\bar{R}) |
\nabla_{\bar{R}}|\psi_n(\bar{R})\rangle \cdot
d\bar{R}.\label{eq:Berrysphase}
\end{equation}
Here $\bar{R}$ are time-dependent parameters of the Hamiltonian, and
the geometric phase becomes an integral in the space of these
parameters (See Eq. (\ref{eq:Berrystirap}) below for a specific
example).

\subsection{STIRAP in lambda system}\label{Sec: STIRAP}
\begin{figure}[htbp]
  \centering
  \includegraphics[width=0.5\textwidth]{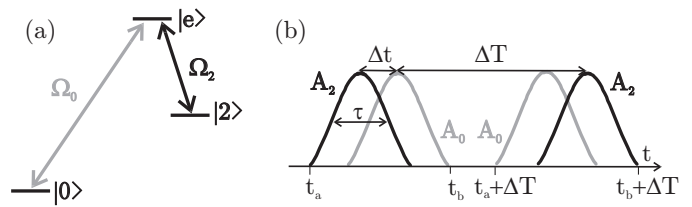}
  \caption{(a) Three-level lambda system with two laser
fields applied with Rabi frequencies $\Omega_0$ and $\Omega_2$. (b)
Pulse sequence consisting of two STIRAP processes separated by
$\Delta T$ in time. The first set of pulses transfers population
from $|0\rangle$ to $|2\rangle$, while the second set transfers it
back. We show the real amplitudes, $A_0$ and $A_2$ of the Rabi
frequencies $\Omega_0$ and $\Omega_2$ as defined in
Eq.~(\ref{eq:omega02}). With this pulse sequence
$\sin\theta={\Omega_0}/({\sqrt{|\Omega_0|^2+|\Omega_2|^2}})$ is zero
before the first pair of pulses arrive, one in between the to
processes and zero after the second pair of pulses. The FWHM of each
pulse is $\tau$ and the delay between pulses within one process is
$\Delta t$.}
  \label{fig:lambda}
\end{figure}

STIRAP is an efficient, adiabatic process for population transfer in
three-level systems. In the method two laser fields are applied to
the atomic lambda system [see Fig. \ref{fig:lambda}(a)]
\cite{bergmann98,unanyan99,gaubatz90,broers92,goldner94,sorensen06,cubel05,lawall94}.
The instantaneous adiabatic dressed states for this system are two
bright states and one dark state. The explicit form of the dark
state $(|D\rangle)$ with zero energy eigenvalue, $\omega^D=0$, reads
\begin{align}\label{eq:darkstirap}
|D\rangle=&\frac{\Omega_2}{\sqrt{|\Omega_0|^2+|\Omega_2|^2}}|0\rangle-\frac{\Omega_0}{\sqrt{|\Omega_0|^2+|\Omega_2|^2}}|2\rangle\\
\notag =&\cos\theta|0\rangle-\sin\theta
e^{i\varphi_2}|2\rangle,\notag
\end{align}
where $\Omega_0$ and $\Omega_2$ are Rabi frequencies (see Fig.
\ref{fig:lambda}), $\tan\theta=|\Omega_0|/|\Omega_2|$ and
$\varphi_2$ is the time-dependent phase difference between the two
laser fields. The real amplitudes of the fields, $A_0(t)$ and
$A_2(t)$ are time-dependent and the Rabi frequencies explicitly read
\begin{align}\label{eq:omega02}
\Omega_{0}(t)&=A_0(t),\\ \notag
\Omega_{2}(t)&=A_2(t)e^{-i\varphi_{2}(t)}.
\end{align}
The pulses are modeled by $\sin^2$-pulses with different amplitudes
\begin{equation}
A_j(t)=\begin{cases}
A_{\textrm{max,j}}\sin^{2}(\frac{\pi (t-t_{sj})}{2 \tau}) & \text{if } t_{sj}<t<t_{sj}+2\tau\\
0 & \text{otherwise}
\end{cases},\label{eq:sinpulses}
\end{equation}
where $t_{sj}$ is the instant of time when the pulse starts and
$\tau$ the FWHM.
With all population initially in the $|0\rangle$-state and only
$\Omega_2$ applied the system is in this dark state. Now,
adiabatically increasing $\Omega_0$ and decreasing $\Omega_2$ causes
population transfer from $|0\rangle$ to $|2\rangle$, while all
population remains in $|D\rangle$ of Eq.~(\ref{eq:darkstirap}). This
implies that the excited state, $|e\rangle$ is never populated and
hence no population is lost due to spontaneous emission.

STIRAP is well-suited for studying geometric phases, because no
dynamic phase is acquired, $\omega^D=0$. The geometric phase is
calculated by Eq.~(\ref{eq:Berrysphase}) with
$\bar{R}=(\theta,\varphi_2)$,
\begin{equation}
\gamma_{D}=i\int_{\bar{R}_i}^{\bar{R}_f}\langle D|
\nabla_{\bar{R}}|D\rangle \cdot d\bar{R}=-\int_{t_i}^{t_f}
\dot{\varphi}_2 \sin^2\theta dt.\label{eq:Berrystirap}
\end{equation}
To obtain a non-zero $\gamma_{D}$, the phase difference between the
two laser fields, $\varphi_2$, should be controlled and have a
time-dependence with non-vanishing $\dot{\varphi}_2$. In order to be
able to control the actual value of $\gamma_{D}$ we require that
$\sin^2\theta=0$ before and after the pulse sequence, ensuring that
only during the sequence a geometric phase is acquired. As stated in
Eq.~(\ref{eq:darkstirap})
$\sin^2\theta=|\Omega_0|^2/(|\Omega_0|^2+|\Omega_2|^2)$, which
implies that during one STIRAP process $\sin^2\theta$ is increased
from zero to one. Applying a second STIRAP process, where $\Omega_0$
is decreased while $\Omega_2$ is increased, decreases $\sin^2\theta$
from one to zero. This second process transfers the population from
$|2\rangle$ back to $|0\rangle$. The whole pulse sequence is shown
in Fig.~\ref{fig:lambda}(b) and after this sequence the system ends
up in
\begin{equation}
|D\rangle=e^{i\gamma_{D}}|0\rangle,
\end{equation}
where $\gamma_{D}$ is calculated form Eq. (\ref{eq:Berrystirap})
with $t_i=t_a$ and $t_f=t_b+\Delta T$ as defined in Fig.
\ref{fig:lambda}(b).

\subsection{STIRAP in tripod system}\label{sec:tripodsystem}
\begin{figure}[htbp]
  \centering
  \includegraphics[width=0.5\textwidth]{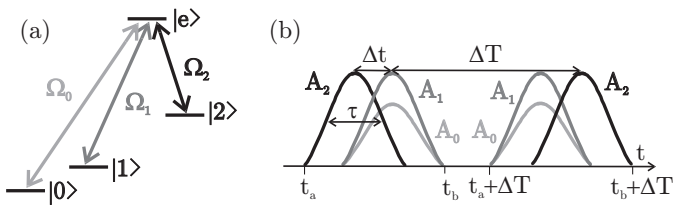}
  \caption{(a) Four-level tripod system with three laser fields applied with Rabi
frequencies $\Omega_0$, $\Omega_1$, $\Omega_2$. (b) Pulse sequence
consisting of two STIRAP processes separated by $\Delta T$ in time.
The first set of pulses transfers population from
$\{|0\rangle,|1\rangle\}$ to $|2\rangle$, while the second set
transfers it back. We show the real amplitudes, $A_0$, $A_1$ and
$A_2$ of the Rabi frequencies $\Omega_0$, $\Omega_1$ and $\Omega_2$
as defined in Eq.~(\ref{eq:omega012}). The FWHM of each pulse is
$\tau$ and the delay between pulses within one process is $\Delta
t$.}
  \label{fig:tripod}
\end{figure}

In the lambda system described above the geometric phase acquired by
the dark state, $|D\rangle$ is a collective phase on the two atomic
states, $|0\rangle$ and $|2\rangle$. We now turn to the tripod
system shown in Fig. \ref{fig:tripod}(a).
This system has the advantage that two dark states are populated and
these acquire different geometric phases leading to relative phases
between the atomic states as well as changes in their population due
to changes in the interference of the two dark states. In the tripod
system we can therefore calculate the geometric phase and see how it
affects the measurable populations and relative phases between the
atomic states. The tripod system further constitute a system, where
a universal set of quantum gates can be implemented (see, e.g.,
\cite{moller07} and references therein). The level structure
consists of three lower states ($|0\rangle$, $|1\rangle$ and
$|2\rangle$) coupled to an excited state, $|e\rangle$, with three
laser fields. These have Rabi frequencies $\Omega_0$, $\Omega_1$ and
$\Omega_2$, respectively, and for simplicity we assume that they are
all on resonance. We restrict ourselves to the case where the
relative phase between $\Omega_0$ and $\Omega_1$, $\varphi_{01}$, is
time-independent while the phase of $\Omega_2$, $\varphi_2(t)$, is
time-dependent. The real amplitudes of all fields, $A_0(t)$,
$A_1(t)$ and $A_2(t)$ are time-dependent and the Rabi frequencies
explicitly read
\begin{align}\label{eq:omega012}
\Omega_{0}(t)&=A_0(t),\\ \notag
\Omega_{1}(t)&=A_1(t)e^{-i\varphi_{01}},\\ \notag
\Omega_{2}(t)&=A_2(t)e^{-i\varphi_{2}(t)}.\notag
\end{align}
The pulses are as in the lambda-system case modeled by the
$\sin^2$-pulses in Eq.~(\ref{eq:sinpulses}). When we apply the pulse
sequence shown in Fig.~\ref{fig:tripod}(b), the STIRAP process
transfers part of the population from $|0\rangle$ and $|1\rangle$ to
$|2\rangle$ and back. During this process geometric phases are
acquired. Calculations of the evolution of the system is presented
in Sec. \ref{sec:closedsystem} for a closed system and in Sec.
\ref{sec:opensystem} for an open system.

\section{Closed system}\label{sec:closedsystem}
In the rotating wave approximation, and with Rabi frequencies as
defined in Eq. (\ref{eq:omega012}), we derive the Hamiltonian
\begin{align}\label{eq:hamiltonnodephasing}
H(t)=\frac{\hbar}{2}\left[\begin{array}{cccc}
0 & 0 & A_{0}(t) & 0\\
0 & 0 & A_{1}(t)e^{i\varphi_{01}} & 0\\
A_{0}(t) & A_{1}(t)e^{-i\varphi_{01}} & 0 &A_{2}(t)e^{-i\varphi_2(t)}\\
0 & 0 & A_{2}(t)e^{i\varphi_2(t)} & 0\\
\end{array}\right]
\end{align}
expressed in the $\{|0\rangle,|1\rangle,|e\rangle,|2\rangle\}$
basis. We parameterize the complex Rabi frequencies as
\begin{align}
\Omega_{0}(t)&=\sin\theta_{01}\sqrt{A_{0}(t)^{2}+A_{1}(t)^{2}},\label{eq:omega0}\\
\Omega_{1}(t)&=\cos\theta_{01}\sqrt{A_{0}(t)^{2}+A_{1}(t)^{2}}e^{-i\varphi_{01}},\label{eq:omega1}\\
\Omega_{2}(t)&=\cos\theta_{H}(t)\sqrt{A_{0}(t)^{2}+A_{1}(t)^{2}+A_{2}(t)^{2}}e^{-i\varphi_2(t)},\label{eq:omega2}
\end{align}
where the two angles are defined as
\begin{align}
\tan\theta_{01}&={A_0(t)}/{A_1(t)},\\
\tan\theta_H(t)&={\sqrt{A_{0}^{2}(t)+A_{1}^{2}(t)}}/{A_2(t)}.
\end{align}
A diagonalization of Eq. (\ref{eq:hamiltonnodephasing}) gives the
four energy eigenvalues
\begin{equation}
\omega^{\pm}=\pm\frac{1}{2}\sqrt{A_{0}^{2}+A_{1}^{2}+A_{2}^{2}}
\hspace{0,2cm}, \qquad\omega^{D_i}=0 \quad(i=1,2),
\end{equation}
and eigenvectors
\begin{widetext}
\begin{align}\label{eq:nodephasingeigenvectors}
|+\rangle&=\frac{1}{\sqrt{2}}\left[\sin\theta_H(t)(\sin\theta_{01}|0\rangle+\cos\theta_{01}e^{i\varphi_{01}}|1\rangle)-|e\rangle+\cos\theta_H(t)e^{i\varphi_2(t)}|2\rangle\right],\\\notag
|-\rangle&=\frac{1}{\sqrt{2}}\left[\sin\theta_H(t)(\sin\theta_{01}|0\rangle+\cos\theta_{01}e^{i\varphi_{01}}|1\rangle)+|e\rangle+\cos\theta_H(t)e^{i\varphi_2(t)}|2\rangle\right],\\\notag
|D_1\rangle&=-\cos\theta_H(t)(\sin\theta_{01}|0\rangle+\cos\theta_{01}e^{i\varphi_{01}}|1\rangle)+\sin\theta_H(t)e^{i\varphi_2(t)}|2\rangle,\\\notag
|D_2\rangle&=\cos\theta_{01}|0\rangle-\sin\theta_{01}e^{i\varphi_{01}}|1\rangle.\\\notag
\end{align}
\end{widetext}
We assume that we start out in a superposition of the two dark
eigenstates at time $t_i$
\begin{equation}
|D(t_i)\rangle=C_{D_1}(t_i)|D_1(t_i)\rangle+C_{D_2}(t_i)|D_2(t_i)\rangle,
\end{equation}
and that the evolution is adiabatic. Then the population stays
within the space spanned by the two dark states and at later times
the wave function is given by
\begin{equation}
|D(t)\rangle=C_{D_1}(t)|D_1(t)\rangle+C_{D_2}(t)|D_2(t)\rangle.\label{eq:darktripod}
\end{equation}
In order to determine the time evolution of the coefficients
$\{C_{D_1}(t),C_{D_2}(t)\}$ we follow Refs.
\cite{unanyan99,wilczek84}. Inserting Eq.~(\ref{eq:darktripod}) into
the time-dependent Schr\"{o}dinger equation yields two coupled
differential equations that can readily be solved and we find the
very simple evolution
\begin{align}
C_{D_1}(t)&=e^{i\gamma_{D_1}}C_{D_1}(t_i), \\
\notag C_{D_2}(t)&=C_{D_2}(t_i). \notag
\end{align}
Here the phase
\begin{equation}
\gamma_{D_1}=-\int_{t_i}^t \dot{\varphi}_2\sin^2\theta_H
dt',\label{eq:closedgammad1}
\end{equation}
acquired by $|D_1\rangle$ is purely geometric because the dark
states do not acquire any dynamic phases, $\omega^{D_i}=0$. Since no
population is transferred between the two dark states the geometric
phase could also be calculated using Eq.~(\ref{eq:Berrysphase}).
This approach was used in \cite{moller07}.

STIRAP (Sec. \ref{Sec: STIRAP}) ensures control of the geometric
phases. The exact pulse sequence is shown in
Fig.~\ref{fig:tripod}(b). The first set of pulses transfers
population partially from $|0\rangle$ and $|1\rangle$ to $|2\rangle$
while the second transfers all population back to $|0\rangle$ and
$|1\rangle$. The amplitudes of $\Omega_0$ and $\Omega_1$ are such
that $\theta_{01}$ is kept constant. The amplitude of $\Omega_2$ is
$A_{\textrm{max},2}=\sqrt{A_{\textrm{max},0}^2+A_{\textrm{max},1}^2}$
and the pulse of $\Omega_2$ is delayed with respect to the pulses of
$\Omega_0$ and $\Omega_1$ and hence $\sin\theta_H(t)$ is varied from
0 to 1 when the first set of pulses arrive, while the second set of
pulses adiabatically turns the $\sin\theta_H(t)$ factor back to 0.
After the whole pulse sequence the system ends up in the final state
\begin{align}
|D(t_f)\rangle=&C_{D_1}(t_i)e^{i\gamma_{D_1}(t_f)}|D_1(t_f)\rangle+C_{D_2}(t_i)|D_2(t_f)\rangle\\\notag
=&[-\sin\theta_{01}C_{D_1}(t_i)e^{i\gamma_{D_1}}+\cos\theta_{01}C_{D_2}(t_i)]|0\rangle\\\notag
&+[-\cos\theta_{01}C_{D_1}(t_i)e^{i\gamma_{D_1}}+\sin\theta_{01}C_{D_2}(t_i)]|1\rangle,\notag
\end{align}
where we have used Eq. (\ref{eq:nodephasingeigenvectors}) in the
second line. In the $\{|0\rangle, |1\rangle\}$-basis, an arbitrary
initial state $|\psi_i\rangle=a_i|0\rangle+b_i|1\rangle$ is
transferred by the unitary matrix $U$ to a final state
$|\psi_f\rangle=U|\psi_i\rangle=a_f|0\rangle+b_f|1\rangle$ with
\begin{widetext}
\begin{equation}
U=\begin{bmatrix}\cos^2\theta_{01}+e^{i\gamma_{D_1}}\sin^2\theta_{01}
& \cos\theta_{01}\sin\theta_{01}
e^{-i\varphi_{01}}(e^{i\gamma_{D_1}}-1)
\\\cos\theta_{01}\sin\theta_{01} e^{i\varphi_{01}}(e^{i\gamma_{D_1}}-1) &
\sin^2\theta_{01}+e^{i\gamma_{D_1}}\cos^2\theta_{01}\end{bmatrix}.
\end{equation} \end{widetext}
By carefully adjusting the amplitudes and phases of the laser
fields, the values of $\theta_{01}$, $\varphi_{01}$ and
$\gamma_{D_1}$ can be controlled and thus generate rotations in the
$\{|0\rangle, |1\rangle\}$-basis. We note that $U$ is the identity
when no geometric phase is acquired, $\gamma_{D_1}=0$. As a special
case $\theta_{01}=\frac{\pi}{8}$,
 $\phi_{01}=\pi$ and $\gamma_{D_1}=-\pi$ implement a Hadamard gate
\begin{equation}
U=\frac{1}{\sqrt{2}}\begin{bmatrix}1 & 1
\\1 & -1\end{bmatrix}.\label{eq:hadamardgate}
\end{equation}
The value $\theta_{01}=\frac{\pi}{8}$ is obtained by choosing
$A_{max,0}=A_{max,1}({\sqrt{2}-1})$,
$A_{max,2}=\sqrt{A^2_{max,0}+A^2_{max,1}}$ and $\varphi_2=t/\tau$,
$\Delta T/\tau=\pi$ assures $\gamma_{D_1}=-\pi$. All simulations
presented throughout this work will use these parameters and the
initial state $|\psi_i\rangle=|0\rangle$ which is transferred to the
final state
$|\psi_f\rangle=U|\psi_i\rangle=(|0\rangle+|1\rangle)/\sqrt{2}$ in
the closed system case.

\section{Open system}\label{sec:opensystem}
We use STIRAP to transfer population among $|0\rangle$, $|1\rangle$
and $|2\rangle$, while keeping the population in the excited state
$|e\rangle$ negligible. Decoherence due to spontaneous emission will
therefore have very little effect, while dephasing caused by, e.g.,
collisions or phase fluctuations of the laser fields, will influence
the evolution. The evolution of the open system can be found by
solving the Lindblad master equation \cite{Haroche06},
\begin{equation}\label{eq:lindblad}
\dot{\rho}=-\frac{i}{\hbar}[H,\rho]-\frac{1}{2}\sum_m(C_m^{\dag}
C_m\rho+\rho C_m^{\dag} C_m)+\sum_m C_m\rho C_m^{\dag},
\end{equation}
where $H$ is the Hamiltonian for the closed system and the
decoherence is described by the Lindblad operators, $C_m$. The
Lindblad master equation results in an ensemble average of the
evolution, but does not reveal a clear distinction between the
geometric and dynamic phases. We wish to follow the evolution of the
wave functions, the acquired geometric and dynamic phases and how
these affect the relative phases between and the population in the
atomic states. Towards this end we use the quantum jump approach,
where the wave function is evolved stochastically
\cite{molmer93,molmer96}. For a small timestep $\Delta t$ the
evolution of the wave function is described as either a jump to
$C_m|\psi(t)\rangle$ or by a no-jump evolution with the
non-Hermitian Hamiltonian $\widetilde{H}=H+H'$, where
$H'=-i\hbar/2\sum_m C_m^{\dag} C_m$. After a timestep with either
the jump or the no-jump evolution the wave function is normalized.
The probability for a jump to $C_m|\psi(t)\rangle$ in $\Delta t$ is
$P_m(t)=\Delta t \langle\psi(t)|C_m^{\dag} C_m|\psi(t)\rangle$. For
the method to be valid the total probability for a jump in $\Delta
t$ has to be small, $P=\sum_m P_m\ll1$. The method leads to many
different traces, which on average reproduce the density matrix. For
further details see, e.g., Refs. \cite{molmer93,molmer96}.

We model the dephasing by a single Lindblad operator,
$C_0=\sqrt{2\Gamma_0}|0\rangle\langle0|$, yielding the master
equation
\begin{align}\label{eq:diffrho}
\dot{\rho}=-\frac{i}{\hbar}[H,\rho]-\left[\begin{array}{cccc}
0 & \Gamma_0\rho_{01} & \Gamma_0\rho_{0e} & \Gamma_0\rho_{02}\\
\Gamma_0\rho_{10} & 0 & 0 & 0\\
\Gamma_0\rho_{e0} & 0 & 0 & 0 \\
\Gamma_0\rho_{20} & 0 & 0 & 0\\
\end{array}\right].
\end{align}
Given $C_0$ we can calculate the no jump evolution with
$\widetilde{H}$ (Sec. \ref{sec:nojump}) as well as the jump traces
(Sec.~\ref{sec:jump}), where the system is projected onto the state
$C_m|\psi(t_j)\rangle\propto|0\rangle$ at the instant of time $t_j$.

\subsection{Non-Hermitian no jump evolution}\label{sec:nojump}
With the Lindblad operator $C_0=\sqrt{2\Gamma_0}|0\rangle\langle0|$,
$H'=-i\hbar\Gamma_0|0\rangle\langle0|$ and the non-Hermitian
Hamiltonian of the tripod system, $\tilde{H}=H+H'$ reads
\begin{small}
\begin{align}\label{eq:hamiltondephasing}
\widetilde{H}(t)=\frac{\hbar}{2}\left[\begin{array}{cccc}
-2i\Gamma_0 & 0 & A_{0}(t) & 0\\
0 & 0 & A_{1}(t)e^{i\varphi_{01}} & 0\\
A_{0}(t) & A_{1}(t)e^{-i\varphi_{01}} & 0 & A_{2}(t)e^{-i\varphi_2(t)} \\
0 & 0 & A_{2}(t)e^{i\varphi_2(t)} & 0\\
\end{array}\right].
\end{align}
\end{small}
In order to determine the time evolution of the system, we transform
into the interaction picture with respect to $H'$
\begin{widetext}
\begin{align}\label{eq:cdiffinteraction}
\widetilde{H}_I(t)=\frac{\hbar}{2}\left[\begin{array}{cccc}
0 & 0 & A_{0}(t)e^{\Gamma_0 (t-t_i)} & 0\\
0 & 0 & A_{1}(t)e^{i\varphi_{01}} & 0\\
A_{0}(t)e^{-\Gamma_0 (t-t_i)} & A_{1}(t)e^{-i\varphi_{01}} & 0 & A_{2}(t)e^{-i\varphi_2(t)} \\
0 & 0 & A_{2}(t)e^{i\varphi_2(t)}& 0
\end{array}\right].
\end{align}
\end{widetext}
The subscript $I$ indicates that the evolution is described in the
interaction picture. This Hamiltonian is non-Hermitian due to the
$\Gamma_0$-exponents and in order to determine the geometric phases
we follow the procedure of Ref. \cite{garrison88}. We diagonalize
$H_I$ and find the eigenvalues
\begin{equation}
\omega_I^{\pm}=\pm\frac{1}{2}\sqrt{A_{0}^{2}+A_{1}^{2}+A_{2}^{2}}
\hspace{0,2cm}, \qquad\omega_I^{D_i}=0 \quad(i=1,2),
\end{equation}
and the right (subscript $r$) and left (subscript $l$) eigenvectors
\begin{widetext}
\begin{align}\label{eq:dephasingeigenvectors}
|+_r\rangle_I&=\frac{1}{\sqrt{2}}\left[\sin\theta_H(t)(\sin\theta_{01}e^{\Gamma_0(t-t_i)}|0\rangle+\cos\theta_{01}e^{i\varphi_{01}}|1\rangle)-|e\rangle+\cos\theta_H(t)e^{i\varphi_2(t)}|2\rangle\right],\\\notag
|-_r\rangle_I&=\frac{1}{\sqrt{2}}\left[\sin\theta_H(t)(\sin\theta_{01}e^{\Gamma_0(t-t_i)}|0\rangle+\cos\theta_{01}e^{i\varphi_{01}}|1\rangle)+|e\rangle+\cos\theta_H(t)e^{i\varphi_2(t)}|2\rangle\right],\\\notag
|D_{1r}\rangle_I&=-\cos\theta_H(t)(\sin\theta_{01}e^{\Gamma_0(t-t_i)}|0\rangle+\cos\theta_{01}e^{i\varphi_{01}}|1\rangle)+\sin\theta_H(t)e^{i\varphi_2(t)}|2\rangle,\\\notag
|D_{2r}\rangle_I&=\cos\theta_{01}e^{\Gamma_0(t-t_i)}|0\rangle-\sin\theta_{01}e^{i\varphi_{01}}|1\rangle,\\\notag
_I\langle+_l|&=\frac{1}{\sqrt{2}}\left[\sin\theta_H(t)(\sin\theta_{01}e^{-\Gamma_0(t-t_i)}\langle0|+\cos\theta_{01}e^{-i\varphi_{01}}\langle1|)-\langle
e|+\cos\theta_H(t)e^{-i\varphi_2(t)}\langle2|\right],\\\notag
_I\langle-_l|_I&=\frac{1}{\sqrt{2}}\left[\sin\theta_H(t)(\sin\theta_{01}e^{-\Gamma_0(t-t_i)}\langle0|+\cos\theta_{01}e^{-i\varphi_{01}}\langle1|)+\langle
e|+\cos\theta_H(t)e^{-i\varphi_2(t)}\langle2|\right],\\\notag
_I\langle
D_{1l}|&=-\cos\theta_H(t)(\sin\theta_{01}e^{-\Gamma_0(t-t_i)}\langle0|+\cos\theta_{01}e^{-i\varphi_{01}}\langle1|)+\sin\theta_H(t)e^{-i\varphi_2(t)}\langle2|,\\\notag
_I\langle
D_{2l}|&=\cos\theta_{01}e^{-\Gamma_0(t-t_i)}\langle0|-\sin\theta_{01}e^{-i\varphi_{01}}\langle1|.\\\notag
\end{align}
\end{widetext}
The left and right eigenvectors fulfill the biorthonormal condition
$\langle i_l|j_r\rangle_I=\delta_{i,j}$ \cite{Faisal87}.

\begin{figure}[htbp]
  \centering
  \includegraphics[width=0.5\textwidth]{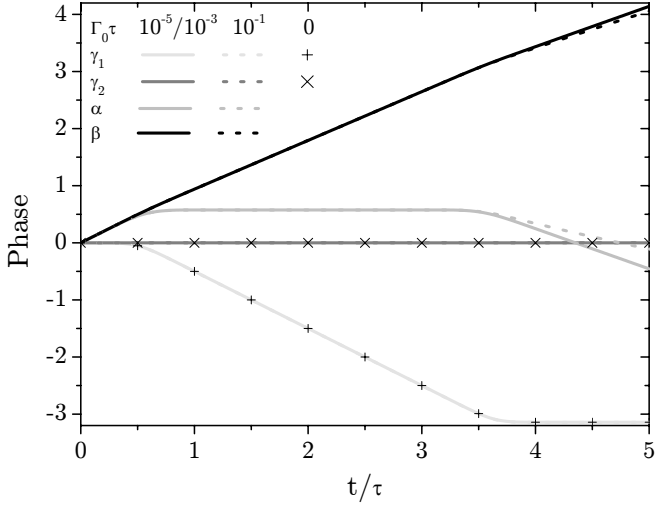}
  \caption{Time evolution of $\gamma_1$, $\gamma_2$, $\alpha$ and $\beta$ for different dephasing rates,
  $\Gamma_0$. The calculations were made with $\sin^2$ pulses (Eq.
(\ref{eq:sinpulses})) and all parameters are given in units of the
pulse width, $\tau$: $\varphi_2=t/\tau$, $A_{max,0}\tau/2\pi=300$,
$A_{max,1}={A_{max,0}}/(\sqrt{2}-1)$,
$A_{max,2}=\sqrt{A^2_{max,0}+A^2_{max,1}}$, $\Delta t/\tau=1$, and
$\Delta T/\tau=\pi$. These parameters lead to
$\theta_{01}=\frac{\pi}{8}$, $\phi_{01}=\pi$ and $\gamma_{1}=-\pi$.
Dephasing rates $\Gamma_0\tau=10^{-5}$ and $\Gamma_0\tau=10^{-3}$
(solid) yield identical results, while $\Gamma_0\tau=10^{-1}$
(dotted) shows a small deviation at $t/\tau\gtrsim3.5$.}
  \label{fig:phases}
\end{figure}

Initially $(t=t_i)$ the eigenvectors of the open system
[Eq.~(\ref{eq:dephasingeigenvectors})] coincide with the
eigenvectors of the closed system
[Eq.~(\ref{eq:nodephasingeigenvectors})], and hence the initial
state of the closed system is also the initial state of the open
system
\begin{equation}
|\psi_i\rangle_I=C_{D_1}(t_i)|D_{1r}(t_i)\rangle_I+C_{D_2}(t_i)|D_{2r}(t_i)\rangle_I.\label{eq:darktripodint}
\end{equation}
The adiabatic STIRAP evolution ensures that the system remains
within the subspace spanned by $\{|D_{1r}(t)\rangle_I,
|D_{2r}(t)\rangle_I\}$. Inserting Eq.~(\ref{eq:darktripodint}) into
the time-dependent Schr\"{o}dinger equation gives a set of equations
that can be solved numerically for $C_{D_1}$ and $ C_{D_2}$. Without
loss of generality we write the solutions as
\begin{align}
C_{D_1}(t)=&e^{-\Gamma_0\alpha(t)}e^{i\gamma_1(t)}C_{D_1}(t_i),\\\notag
C_{D_2}(t)=&e^{-\Gamma_0\beta(t)}e^{i\gamma_2(t)}C_{D_2}(t_i),\notag
\end{align}
and expand the wave function as
\begin{align}
|\psi(t)\rangle_I=&e^{-\Gamma_0\alpha(t)}e^{i\gamma_1(t)}C_{D_1}(t_i)|D_{1r}(t)\rangle_I\\\notag
+&e^{-\Gamma_0\beta(t)}e^{i\gamma_2(t)}C_{D_2}(t_i)|D_{2r}(t)\rangle_I.\notag
\end{align}
The two dark states each acquire a complex geometric phase composed
by real ($\gamma_1$ and $\gamma_2$) and imaginary parts
($\Gamma_0\alpha(t)$ and $\Gamma_0\beta(t)$) parameterized by the
dephasing rate, $\Gamma_0$. As an example we choose an initial state
$|\psi_i\rangle=|0\rangle$ and apply the pulse sequence in Fig.
\ref{fig:tripod}(b) with parameters leading to
$\theta_{01}=\frac{\pi}{8}$, $\phi_{01}=\pi$ and $\gamma_{D_1}=-\pi$
(see details in caption of Fig.~\ref{fig:phases}). As discussed in
Sec.~\ref{sec:closedsystem} this evolution corresponds to the
Hadamard gate, Eq.~(\ref{eq:hadamardgate}) for the closed system. In
Fig.~\ref{fig:phases} we show the evolution of the phases as a
function of time for different values of the dephasing rate
$\Gamma_0$. The real geometric phases $\gamma_1$ (light grey) and
$\gamma_2$ (grey) are unaffected by the dephasing and their values
are identical to the analytical result for the closed system [Eq.
(\ref{eq:closedgammad1})], which are marked with crosses in Fig.
\ref{fig:phases}. The exponents $\alpha$ and $\beta$ are also almost
unaffected by the dephasing rate. The results for
$\Gamma_0\tau=10^{-5}$ and $\Gamma_0\tau=10^{-3}$ (solid) are
identical while increasing the dephasing to $\Gamma_0\tau=10^{-1}$
(dotted) gives only a small deviation. Experimentally dephasing
rates can be kept below $\Gamma_0\tau=10^{-3}$ and for these values
of $\Gamma_0$, $\alpha$ and $\beta$ are unaffected by the dephasing
rate and hence the imaginary part of the phases $\Gamma_0\alpha$ and
$\Gamma_0\beta$ scales linear with the dephasing rate. It should be
noted that the value of $\alpha$ and $\beta$ depend on the initial
state while $\gamma_1$ and $\gamma_2$ are unaffected and equal to
the values in the closed system case.
Going back to the Schr\"{o}dinger picture yields
\begin{align}
|\psi(t)\rangle=\frac{1}{\sqrt{N}}e^{-\Gamma_0(t-t_i)|0\rangle\langle0|}[&e^{-\Gamma_0\alpha}e^{i\gamma_1}C_{D_1}(t_i)|D_{1r}\rangle_I\\\notag
+&e^{-\Gamma_0\beta}e^{i\gamma_2}C_{D_2}(t_i)|D_{2r}\rangle_I],\notag
\end{align}
where the wave function was re-normalized (factor $1/\sqrt{N}$)
because the non-Hermitian Hamiltonian does not preserve the norm. In
the $\{|0\rangle, |1\rangle\}$-basis an initial state in the
Schr\"{o}dinger picture $|\psi_i\rangle=a_i|0\rangle+b_i|1\rangle$
is transferred to a final non-normalized state
$|\psi_f\rangle_{non-norm}=L|\psi_i\rangle=a_f|0\rangle+b_f|1\rangle$,
with
\begin{widetext}
\begin{equation}
L=\begin{bmatrix}e^{i\gamma_{2}}\cos^2\theta_{01}e^{-\Gamma_0\beta}+e^{i\gamma_{1}}\sin^2\theta_{01}e^{-\Gamma_0\alpha}
& \cos\theta_{01}\sin\theta_{01}
e^{-i\varphi_{01}}(e^{i\gamma_{1}}e^{-\Gamma_0\alpha}-e^{i\gamma_{2}}e^{-\Gamma_0\beta})
\\\cos\theta_{01}\sin\theta_{01} e^{i\varphi_{01}}(e^{i\gamma_{1}}e^{-\Gamma_0\alpha}-e^{i\gamma_{2}}e^{-\Gamma_0\beta}) &
e^{i\gamma_{2}}\sin^2\theta_{01}e^{-\Gamma_0\beta}+e^{i\gamma_{1}}\cos^2\theta_{01}e^{-\Gamma_0\alpha}\end{bmatrix}.\label{eq:nojumpmatrix}
\end{equation} \end{widetext}
The normalization constant,
$N=|C_{D_1}(t_i)|^2e^{-2\Gamma_0\alpha}+|C_{D_2}(t_i)|^2e^{-2\Gamma_0\beta}$,
depends on the initial state, and the final state reads
$|\psi_f\rangle=\frac{1}{\sqrt{N}}|\psi_f\rangle_{non-norm}$. While
the phases $\gamma_{1}$ and $\gamma_{2}$ are robust with respect to
dephasing, the appearance of the imaginary phases, affects the
population of the states.
\begin{figure}[htbp]
  \centering
  \includegraphics[width=0.45\textwidth]{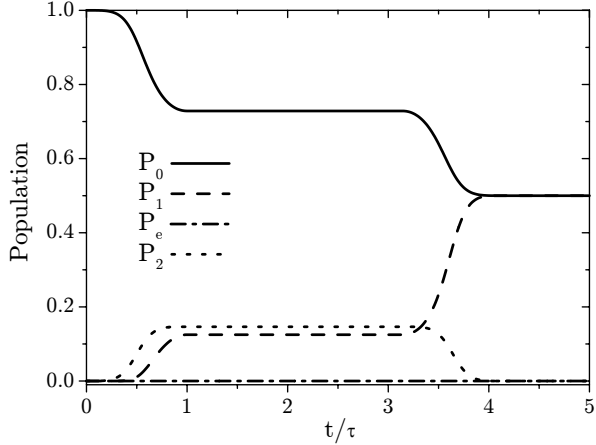}
  \caption{Evolution of the population in the $|0\rangle$-state ($P_0$, solid), the $|1\rangle$-state ($P_1$, dashed),the $|e\rangle$-state ($P_e$, dashed-dotted) and the $|2\rangle$-state ($P_2$, dotted) with all population initially in the $|0\rangle$-state.
  The population evolve to $P_0=1/2$ and $P_1=1/2$ when no dephasing
  is present. With dephasing ($\Gamma_0T_0=10^{-3}$) the
  master equation as well as the no jump evolution
  [Eq.~(\ref{eq:nojumpmatrix})] show no deviation from the no
  dephasing case on the scale of this figure. The calculations were made
  with $\sin^2$ pulses [Eq.~(\ref{eq:sinpulses})] and all parameters are given in units of
the pulse width, $\tau$: $\varphi_2=t/\tau$,
$A_{max,0}\tau/2\pi=300$, $A_{max,1}=A_{max,0}/(\sqrt{2}-1)$,
$A_{max,2}=\sqrt{A^2_{max,0}+A^2_{max,1}}$, $\Delta t/\tau=1$, and
$\Delta T/\tau=\pi$. These parameters lead to
$\theta_{01}=\frac{\pi}{8}$, $\phi_{01}=\pi$ and $\gamma_{1}=-\pi$.}
  \label{fig:sml}
\end{figure}
Figure~\ref{fig:sml} shows the evolution of the population of the
four states $(|0\rangle,|1\rangle,|e\rangle,|2\rangle)$ with initial
state $|\psi_i\rangle=|0\rangle$ and the pulse sequence in
Fig.~\ref{fig:tripod}(b) with parameters $\theta_{01}=\pi/8$,
$\varphi_{01}=\pi$ and $\gamma_{D_1}=-\pi$ (see detailed parameters
in the caption of Fig.~\ref{fig:sml}). In the closed system these
parameters lead to an implementation of the Hadamard gate and hence
final populations $P_0=\frac{1}{2}$ and $P_1=\frac{1}{2}$. When we
introduce dephasing $|\psi_i\rangle=|0\rangle$ leads to final
population found from the no-jump evolution
[Eq.~(\ref{eq:nojumpmatrix})]
\begin{align}
P_0=&\frac{(\cos^2\theta_{01}e^{-\Gamma_0\beta}-\sin^2\theta_{01}e^{-\Gamma_0\alpha})^2}{\sin^2\theta_{01}e^{-2\Gamma_0\alpha}-\cos^2\theta_{01}e^{-2\Gamma_0\beta}},\\
\notag
P_1=&\frac{(\sin\theta_{01}\sin\theta_{01}(e^{-\Gamma_0\beta}+e^{-\Gamma_0\alpha}))^2}{\sin^2\theta_{01}e^{-2\Gamma_0\alpha}-\cos^2\theta_{01}e^{-2\Gamma_0\beta}}.
\notag
\end{align}
The deviation from the closed system case
$(P_0=(\cos^2\theta_{01}-\sin^2\theta_{01})^2,P_1=4\sin^2\theta_{01}\cos^2\theta_{01})$
is thus determined by the the two imaginary geometric phases,
$\Gamma_0\alpha$ and $\Gamma_0\beta$. The values of $\alpha$ and
$\beta$ are found by numerically solving the time-dependent
Schr\"{o}dinger equation. On the scale of Fig.~\ref{fig:sml} there
is no deviation between the closed and the open system results for
realistic dephasing rates.

\subsection{Jump evolution}\label{sec:jump}
\begin{figure}[htbp]
  \centering
  \includegraphics[width=0.5\textwidth]{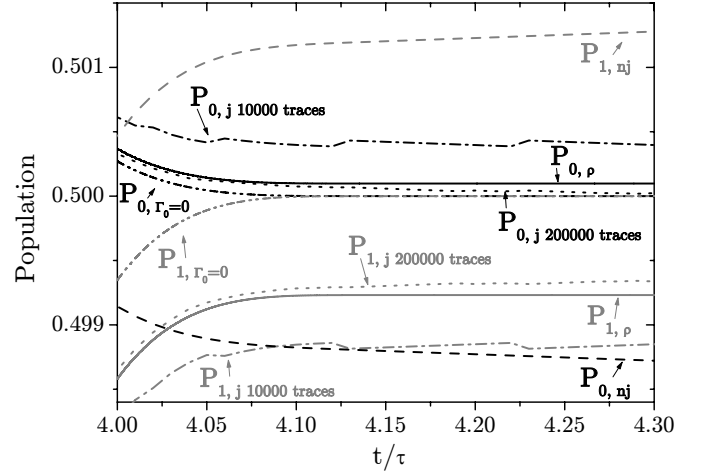}
  \caption{Population in the $|0\rangle$-state ($P_0$,
  black), the $|1\rangle$-state ($P_1$,
  grey), when the system is not subject to dephasing ($P_{i,\Gamma_0=0}$, dot-dot-dash)
  as well a when dephasing ($\Gamma_0\tau=10^{-3}$) is present. The graph shows only the final part of the
  time evolution, where the differences can be distinguished. With dephasing the
  full curve ($P_{i,\rho}$) shows the solution of the master equation [Eq.~(\ref{eq:lindblad})], the dashed curve ($P_{i,nj}$) the no-jump
  trace, while the last two curves are averages over $10000$
  ($P_{i,j\textrm{ 10000 traces}}$ dash-dot-dash) and $200000$ ($P_{i,j\textrm{ 200000 traces}}$, dotted) jump and no-jump
  traces. The calculations were made with $\sin^2$ pulses (Eq.
(\ref{eq:sinpulses})) and all parameters are given in units of the
pulse width, $\tau$: $\varphi_2=t/\tau$, $A_{max,0}\tau/2\pi=300$,
$A_{max,1}=A_{max,0}/(\sqrt{2}-1)$,
$A_{max,2}=\sqrt{A^2_{max,0}+A^2_{max,1}}$, $\Delta t/\tau=1$, and
$\Delta T/\tau=\pi$.}
  \label{fig:rhojump}
\end{figure}

For the present choice of parameters the system jumps in a small
part of the Monte Carlo traces to the $|0\rangle$-state due to
$C_0=\sqrt{2\Gamma_0}|0\rangle\langle0|$. If a jump occurs at $t_j$
we expand the wave function ($|0\rangle$) in the adiabatic basis of
the instantaneous eigenstates
\begin{align}
|\psi(t_j)\rangle=&|0\rangle\\ \notag
=&-\cos\theta_H(t_j)\sin\theta_{01}|D_{1r}(t_j)\rangle+\cos\theta_{01}|D_{2r}(t_j)\rangle\\\notag
&+\frac{1}{\sqrt{2}}
\sin\theta_H(t_j)\sin\theta_{01}(|+_r(t_j)\rangle+|-_r(t_j)\rangle).\notag
\end{align}
After the jump the evolution is described by the no-jump
non-Hermitian Hamiltonian [Eq.~(\ref{eq:cdiffinteraction})]. In the
adiabatic basis we can describe this evolution by calculating the
geometric and dynamic phases acquired by the four states. The dark
part of the wave function evolves as described for the no-jump
evolution. The bright eigenstates $(|+_r\rangle,|-_r\rangle)$ are
separated energetically from each other and from the dark states,
such that there is no diabatic population transfer among these. The
bright states acquire a dynamic phase as well as a complex geometric
phase that can be calculated directly Eq.~(\ref{eq:Berrysphase}).
The complex geometric phase is the same for the two bright states
\begin{align}
\gamma_B=&i\int_{\bar{R}_i}^{\bar{R}_f}\langle\pm_l|
\nabla_{\bar{R}}|\pm_r\rangle \cdot d\bar{R}\\ \notag
=&\frac{1}{2}i\Gamma_0\sin\theta^2_{01}\int_{t_j}^t\sin^2\theta_Hdt'-\frac{1}{2}\int_{t_j}^t\dot{\varphi}_2\cos^2\theta_H
dt'\\\notag \equiv&i\Gamma_0\delta+\gamma_b\notag
\end{align}
while the dynamic phase is different for the
$(|+_r\rangle,|-_r\rangle)$-states
\begin{equation}
\vartheta_{\pm}=\mp\frac{1}{2}\int_{t_j}^t
\sqrt{A_0^2+A_1^2+A_2^2}dt'.\notag
\end{equation}
The wave function at later times in the interaction picture is
\begin{align}
|\psi(t)\rangle=&-\cos\theta_H(t_j)\sin\theta_{01}e^{-\Gamma_0\alpha}e^{i\gamma_1}|D_{1r}\rangle\\\notag&+\cos\theta_{01}e^{-\Gamma_0\beta}e^{i\gamma_2}|D_{2r}\rangle\\\notag
&+\frac{1}{\sqrt{2}}\sin\theta_H(t_j)\sin\theta_{01}e^{-\Gamma_0\delta}e^{i\gamma_b}e^{i\vartheta_+}|+_r\rangle\\\notag
&+\frac{1}{\sqrt{2}}\sin\theta_H(t_j)\sin\theta_{01}e^{-\Gamma_0\delta}e^{i\gamma_b}e^{i\vartheta_-}|-_r\rangle.\notag
\end{align}
Going back to the Schr\"{o}dinger picture and calculating the final
populations in the atomic states yields
\begin{align}
P_0=&\frac{1}{N}|\cos^2\theta_{01}e^{-\Gamma_0\beta}-\sin^2\theta_{01}\cos\theta_{H}(t_j)e^{-\Gamma_0\alpha}e^{i\gamma_1})|^2,\\
\notag
P_1=&\frac{1}{N}\sin^2\theta_{01}\cos^2\theta_{01}|e^{-\Gamma_0\beta}+\cos\theta_{H}(t_j)e^{-\Gamma_0\alpha}e^{i\gamma_1}|^2,
\\\notag
P_e=&\frac{1}{N}\sin^2\theta_{01}\sin^2\theta_{H}(t_j)e^{-2\Gamma_0\delta}\sin^2{\vartheta_-},\\\notag
P_2=&\frac{1}{N}\sin^2\theta_{01}\sin^2\theta_{H}(t_j)e^{-2\Gamma_0\delta}\cos^2{\vartheta_-},\notag
\end{align}
where the normalization constant is given as
$N=\sin^2\theta_{01}\cos^2\theta_{H}(t_j)e^{-2\Gamma_0\alpha}+\cos^2\theta_{01}e^{-2\Gamma_0\beta}+\sin^2\theta_{01}\sin^2\theta_{H}(t_j)e^{-2\Gamma_0\delta}$.
Phases are only acquired from the latest jump time, $t_j$. The final
populations can in this way be calculated after each Monte Carlo
trace and the deviations from the closed system case can be
explained by the complex geometric phases acquired by the adiabatic
states. When all complex geometric phases are carefully taken into
account an average over many traces reproduces the numerical
solution of the master equation [Eq.~(\ref{eq:diffrho})]. This is
shown in Fig.~\ref{fig:rhojump}, where we enlarge the last part of
the evolution of the populations shown in Fig.~\ref{fig:sml}. On
this scale deviations between the closed and the open system are
visible. The closed system (dash-dot-dot curves) yields final
populations $P_0=1/2$ and $P_1=1/2$ as expected for the initial
state $|\psi_i\rangle=|0\rangle$ and the Hadamard gate applied
(parameters are specified in the caption of Fig.~\ref{fig:rhojump}).
The evolution of the open system is determined either by solving the
master equation (solid curves) or by the Monte Carlo method. The
final populations predicted by the no-jump evolution (dashed curves)
deviates from the master equation (solid curves) on the order of
$10^{-3}$. An average over $10000$ Monte Carlo traces
(dash-dot-dash) reduces the deviation to an order of $10^{-4}$,
while averaging over $200000$ traces (dotted curves) reduces it
further to an order of $10^{-5}$.

\subsection{Fidelity of Hadamard gate}
\begin{figure}[htbp]
  \centering
  \includegraphics[width=0.5\textwidth]{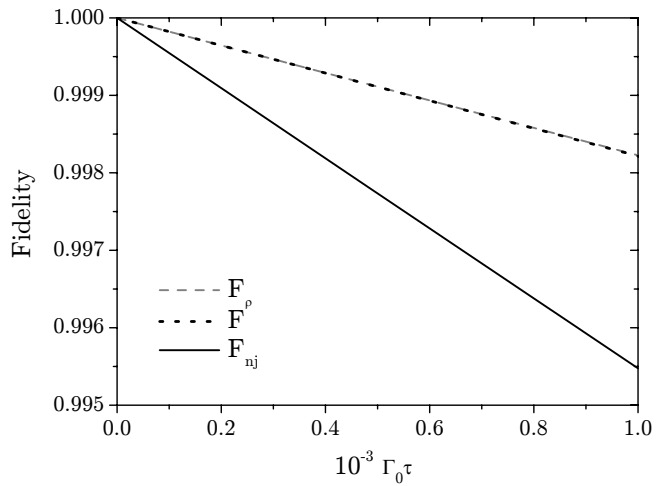}
  \caption{Fidelity as a function of the dephasing rate, $\Gamma_0$. The dashed grey curve shows the density matrix calculation, $F_{\rho}$ [Eq.
  (\ref{eq:fidelityrho})], the dotted black curve the Monte Carlo
  calculation including traces with no jump as well as one jump ($F$) and solid black includes only no-jump traces ($F_{nj}$) [Eq.~\ref{eq:fidelitetnojump}]. The calculations were made
with $\sin^2$ pulses [Eq.~(\ref{eq:sinpulses})] and all parameters
are given in units of the pulse width, $\tau$: $\varphi_2=t/\tau$,
$A_{max,0}\tau/2\pi=300$, $A_{max,1}=A_{max,0}/(\sqrt{2}-1)$,
$A_{max,2}=\sqrt{A^2_{max,0}+A^2_{max,1}}$, $\Delta t/\tau=1$, and
$\Delta T/\tau=\pi$.}
  \label{fig:fidelity}
\end{figure}

The calculated Monte Carlo traces can be used to determine the
fidelity of the Hadamard gate. For a given initial state
$|\psi_i\rangle$ we can determine the fidelity as the overlap
between the target (closed-system) wave function $|\psi_0\rangle$
and the final Monte Carlo wave functions (no jump
$|\psi_{nj}\rangle$ or one jump at $t_j$ $|\psi_j(t_j)\rangle$)
weighed by the probability of each trace (no jump $P_{nj}$ or one
jump $P_j(t_j)$). The contributions to the fidelity from traces with
more than one jump ($F_{i,mj}$) is negligible for realistic
dephasing rates. The fidelity, accordingly reads
\begin{align}
F_i=&P_{nj}|\langle \psi_0 |\psi_{nj}\rangle|^2+\int
P_j(t_j)|\langle\psi_0 |\psi_j(t_j)\rangle|^2dt_j+F_{i,mj}\\\notag
=&|\langle \psi_0 |\psi_{nj,non-norm.}\rangle|^2\\\notag &+ \int
|\langle \psi_0 |\psi_{j,non-norm.}(t_j)\rangle|^2dt_j+F_{i,mj},
\end{align}
where we use the non-normalized Monte Carlo wave functions to avoid
calculating the probability distributions directly. As a first
approximation the fidelity can be calculated neglecting the jump
traces
\begin{align}\label{eq:fidelitetnojump}
F_{i,nj}=&|\langle \psi_0 |\psi_{nj,non-norm.}\rangle|^2=|\langle
\psi_iU_0^\dagger|L\psi_i\rangle|^2\\\notag
=&e^{-2\Gamma_0\alpha_{nj}}|C_{D_1}(t_i)|^4+e^{-2\Gamma_0\beta_{nj}}|C_{D_2}(t_i)|^4\\\notag
&+2e^{-\Gamma_0(\alpha_{nj}+\beta_{nj})}|C_{D_1}(t_i)|^2|C_{D_2}(t_i)|^2\\\notag
&\times\cos(\gamma_{1,nj}-\gamma_{2,nj}-\gamma_{D_1}).
\end{align}
The decrease in the fidelity is governed by the geometric phases
acquired during the no-jump evolution. The non-normalized wave
function for traces with one jump at the instant $t_j$ is found in
three steps. The system evolves under the non-Hermitian Hamiltonian
until $t_j$, $|\psi_j(t_j)\rangle=L|\psi_j(t_i)\rangle$, where it
acquires geometric phases $-\Gamma_0\alpha+i\gamma_1$ and
$-\Gamma_0\beta+i\gamma_2$. At $t_j$ the system jumps and
$|\psi_j(t_j)\rangle=\sqrt{2\Gamma_0}|0\rangle\langle0|\psi_j(t_j)\rangle$.
Finally the system evolves under the non-Hermitian Hamiltonian from
$t_j$ to $t_f$, $|\psi_j(t_f)\rangle=L|\psi_j(t_j)\rangle$, where it
acquires geometric phases $-\Gamma_0\alpha'+i\gamma'_1$ and
$-\Gamma'_0\beta+i\gamma'_2$. The fidelity from the jump traces will
hence be proportional to $\sqrt{2\Gamma_0}^2$
\begin{equation}
F_i=F_{i,nj}+2\Gamma_0\int\xi_{i,j}(\alpha,\gamma_1,\alpha',\gamma'_1,\beta,\gamma_2,\beta',\gamma'_2)
dt_j,
\end{equation}
where
$\xi_{i,j}(\alpha,\gamma_1,\alpha',\gamma'_1,\beta,\gamma_2,\beta',\gamma'_2)$
is determined by the geometric phases in the jump traces with jump
at $t_j$. Traces with more than one jump will contribute with terms
proportional to higher orders of $\Gamma_0$ and are therefore
neglected. The fidelity determined from the Monte Carlo traces can
be compared with the Uhlmann state fidelity calculated from the
final closed system density matrix $\rho_0$ and the final open
system density matrix $\rho_{(\Gamma_0)}$ \cite{uhlmann76}
\begin{equation}\label{eq:fidelityrho}
F_{\rho,i}=\left(\textrm{Tr}\sqrt{\rho_0^{1/2}\rho_{(\Gamma_0)}\rho_0^{1/2}}\right)^2.
\end{equation}
Both $F_i$ and $F_{\rho,i}$ give the fidelity for a given initial
state. The average fidelity can be found by integrating over the
surface of the Bloch sphere
\begin{equation}
F=\frac{1}{4\pi}\int F_i d\Omega.
\end{equation}
This averaging procedure can be simplified to only averaging over
the six axial pure initial states on the Bloch sphere,
$\Lambda=$\{\mbox{$|0\rangle$}, \mbox{$|1\rangle$},
\mbox{$\frac{1}{\sqrt{2}}(|0\rangle+|1\rangle)$},
\mbox{$\frac{1}{\sqrt{2}}(|0\rangle-|1\rangle)$},
\mbox{$\frac{1}{\sqrt{2}}(|0\rangle+i|1\rangle)$},
\mbox{$\frac{1}{\sqrt{2}}(|0\rangle-i|1\rangle)$}\}\cite{bowdrey02}
\begin{equation}
F=\frac{1}{6}\sum_{|\psi_i\rangle\epsilon\Lambda}F_i.\\
\end{equation}
In Fig.~\ref{fig:fidelity} we show the average fidelity as a
function of the dephasing rate by the Monte Carlo method when only
no-jump traces are taken into account (solid, black curve) and when
traces with no or one jump are included (dotted, black curve). These
are compared with the full master equation solution (dashed, grey
curve). The fidelities are all calculated for the Hadamard gate
implemented by the parameters $\theta_{01}=\pi/8$,
$\varphi_{01}=\pi$ and $\gamma_{D_1}=-\pi$ as in all previous
numerical results.
The fidelity decreases as expected when the dephasing rate
increases, but is still acceptable when the system is subject to
realistic dephasing rates. The no-jump results gives a lower bound
on the fidelity but it is necessary to include traces with one jump
in order to get a satisfactory accuracy for realistic dephasing
rates. Eq.~(\ref{eq:fidelitetnojump}) shows explicitly how the
no-jump fidelity depends on the geometric phases acquired during the
no-jump evolution.

\section{Summary and conclusion}\label{sec:conclusion}
We presented a method to describe the adiabatic evolution of an open
system using the quantum Monte Carlo method and keeping track of all
acquired phases. This method has the advantage that it reveals the
evolution of the single quantum trajectories and discloses the
geometric or dynamic nature of the acquired phases in the adiabatic
basis. We considered a tripod system with three laser fields applied
and calculated its instantaneous adiabatic eigenstates. The tripod
system is subject to a double STIRAP process and during this the
adiabatic eigenstates acquire complex geometric phases. In the
closed system case the geometric phases are purely real
($\gamma_{D_1}$) and all population is at all times in the two dark
(with zero energy eigenvalues) adiabatic eigenstates. The acquired
geometric phases create a phase difference between the two dark
states and hence generate a rotation in the atomic
$\{|0\rangle,|1\rangle\}$-basis. With the right parameter choice
this rotation implements the Hadamard gate, which we have used as an
example in the numerical simulations.

When dephasing is present we used the quantum Monte Carlo method,
where the system either follow a non-Hermitian no-jump evolution
during the whole time sequence or at one or more instants of time
jump to the $|0\rangle$-state. Before, in between and after jumps
the system follows the non-Hermitian no-jump evolution. During the
non-Hermitian evolution the instantaneous adiabatic eigenstates
acquire complex geometric phases. These deviate from the closed
system case mainly because they contain non-negligible imaginary
parts, which lead to a decay of the adiabatic eigenstates and hence
influence the populations resulting in imperfect gate performance.
The specific Hadamard gate simulations show that the fidelity is
still appreciable at realistic dephasing rates.

\begin{acknowledgments}
This work is supported by the Danish Research Agency (Grant. No.
2117-05-0081).
\end{acknowledgments}

\end{document}